# A Tripartite Framework for Understanding the Embodied Mind: A Comprehensive Review of Bodily Map Study

Tatsuya Daikoku, Maiko Minatoya, Masaki Tanaka

Graduate School of Information Science and Technology, The University of Tokyo, Tokyo, Japan

*** Corresponding author**
Tatsuya Daikoku
Graduate School of Information Science and Technology, The University of Tokyo, Tokyo, Japan,
7-3-1 Hongo, Bunkyo-ku, 113-8656, Tokyo, Japan
Email: daikoku.tatsuya@mail.u-tokyo.ac.jp

# Abstract

People often speak of being “moved beyond words” when witnessing awe-inspiring natural beauty, experiencing profound moral elevation, or encountering powerful works of art and music. Such moments evoke intensely embodied sensations—a swelling in the chest, a quickened heartbeat, or a sudden stillness that floods the entire body. These experiences are difficult to articulate, yet they are vividly real and spatially anchored in bodily experience. How, then, can science begin to access and understand these deeply felt, yet ineffable, emotional states?

This review examines the emerging method of body mapping, which captures how individuals subjectively localize emotional sensations across the body. Drawing on over two dozen empirical studies, we integrate findings from affective neuroscience, psychophysiology, cognitive linguistics, and cross-cultural psychology to propose a tripartite framework for understanding the origins of emotional bodily maps: (1) bottom-up physiological signals, (2) top-down behavioral engagement (including motor actions), and (3) conceptual and metaphorical constructions. From racing hearts and heavy limbs to sensations of insight in the gut, we argue that bodily maps emerge from a dynamic integration of physiological signals, action tendencies, and symbolic conceptual schemas.

Crucially, we argue that bodily maps are not mere reflections of physiological monitoring, but expressive interfaces between brain, body, and conceptualization. Because they rely on introspective sensation rather than verbal articulation, bodily maps offer a promising language-independent tool for assessing emotional experience—particularly in cross-cultural research, developmental studies, and clinical contexts where verbal reporting may be limited or unreliable. Ultimately, bodily maps reveal the hidden geography of emotion—those feelings that elude precise definition or verbalization, like being moved beyond words, yet are undeniably real and embodied—offering a compelling window into how emotions are formed, shared, and lived through the body.

# 1. Introduction

Emotions are deeply grounded in embodied experience. A growing body of research highlights the central role of interoception—the brain's capacity to monitor internal physiological states such as heart rate, respiration, and gastrointestinal activity—in shaping emotional experience (Critchley & Harrison, 2013; Barrett & Simmons, 2015). This perspective suggests that emotions arise from the brain's interpretation of physiological signals. However, the link between physiological state and emotional category is far from one-to-one. Autonomic measures such as heart rate, blood pressure, and skin conductance show substantial overlap across emotions, challenging the notion of discrete physiological "fingerprints" for specific and finely differentiated affective states in humans (Siegel et al., 2018). Nonetheless, individuals consistently report localized bodily sensations in association with emotional experiences—such as a fluttering in the stomach during anxiety, tightness in the chest with sadness, or a flush of heat rising to the face in anger.

Such embodied experiences are deeply embedded in language and metaphor. Expressions like "butterflies in the stomach," "cold feet," or "losing one's head" reflect an intuitive mapping between emotional states and body regions. These metaphors are not mere figures of speech—they encode culturally meaningful ways of experiencing and communicating affect. Empirical findings support this intuition: emotional episodes often elicit vivid interoceptive sensations—goosebumps, a racing heartbeat, or bodily heaviness—that are phenomenologically rich and play functional roles in emotional regulation and behavior (Craig, 2002; Critchley & Harrison, 2013).

A method for investigating the bodily grounding of emotion is the use of bodily maps. It is important to clarify that the term body mapping as used here, in principle, does not refer to somatotopic mapping in the brain (e.g., cortical homunculi in somatosensory cortex), but rather to subjective reports of where emotions are felt within the body. These maps capture phenomenological associations between emotional states and bodily regions, which may be shaped by interoceptive signals, behavioral tendencies, and cultural metaphors. While this focus is more limited than neural somatotopy, it offers complementary insights into how people consciously experience and communicate emotions through their bodies. Their findings revealed that distinct emotions produce consistent and spatially differentiable patterns of activation and deactivation across the body. For instance, negative emotions such as anger and fear predominantly activate the upper torso and head, while positive emotions like happiness and love are associated with widespread activation. These maps were strikingly consistent across Western and East Asian cultures (Nummenmaa et al., 2014), suggesting that emotional topographies may reflect universal bodily representations of affect. Building on this foundation, recent studies have explored how more complex or social emotions—such as morality (Atari et al., 2020), love (Rinne et al., 2023), empathy (Sachs et al., 2019) and aesthetic emotion induced by music and arts (Putkinen et al., 2024; Daikoku et al., 2024)—can also be captured through body mapping methods (Table 1).

This growing body of work raises questions about the mechanisms by which emotional experiences become embodied: Are these bodily maps driven by bottom-up physiological responses to stimuli, such as increased heart rate by surprisal stimuli? Or do they arise from top-down behavioral tendencies, like the impulse to dance with excitement? Alternatively, could certain bodily expressions of emotion—particularly those rooted in metaphor, conceptual schemas, or language (e.g., "blow one's head") —emerge independently of any actual physiological change? Despite the expanding use of bodily maps in affective science, the origins and nature of the subjective sensations they represent remain poorly understood. This ambiguity hinders our ability to interpret the psychological and physiological meaning of these maps and limits insight into how and why they differ across individuals, emotional categories, and cultural or situational contexts (Herbert et al., 2011; Pollatos et al., 2009; Lyons et al., 2021; Lloyd et al., 2021). Throughout this review, we use bodily maps to mean graded topographies of felt intensity distributed across the entire body, not a claim that emotions are anatomically localized to isolated parts. These maps summarize relative increases and decreases in sensation and readiness, rather than exclusive loci.

In this review, our primary goal is to address these questions by synthesizing recent findings on bodily maps of emotion, clarifying what is currently known and identifying key gaps in understanding. Building on this foundation, we tentatively introduce a tripartite framework as a heuristic for interpreting the possible origins of emotional bodily maps. This framework is intended to stimulate discussion and generate hypotheses for future research. By situating the framework in this way, the review aims both to consolidate current evidence and to suggest productive directions for investigating how emotions are constructed, experienced, and expressed through the body.

# 2. Mapping Emotion in the Body

Over the past decade, body mapping has emerged as a powerful method for capturing the embodied emotion (Table 1, Figure 1). Unlike verbal reports or physiological recordings, this approach offers a direct, introspective window into how people "subjectively" localize and represent their emotional states in the body (Daikoku et al., 2024). The method typically involves participants indicating regions of increased or decreased sensation on a body silhouette, in response to specific emotional states. While deceptively simple, this technique has proven remarkably robust, revealing rich and structured representations of emotion across populations and contexts (Nummenmaa et al., 2014; Volynets et al., 2020).

We comprehensively reviewed relevant literature by searching PubMed, Web of Science, and Scopus, as well as Google Scholar and Google Search with the keywords "Bodily map" and "Body map." We specifically focused on peer-reviewed studies examining the relationship between music, language, emotions, and bodily sensations, with particular attention to studies employing the body silhouette method. The initial search and screening were conducted by T.D., and the selection was carefully reviewed and checked for completeness by co-authors M.T. and M.M. We identified two principal variants of the silhouette method. The first, developed by Nummenmaa et al. (2014), in-volved presenting participants with a stimulus (e.g., music clip) and a blank human silhouette; participants were asked to color the regions of the body where they felt activity changes while listening. The second, introduced by Daikoku et al. (2024), was a click-based method: while or im-mediately after stimulus presentation, participants were asked to respond within 10 seconds by clicking on the body image to indicate where they felt sensations from the chords.

Collective findings from over two dozen body mapping studies highlight three key regularities. First, basic emotions such as anger, fear, sadness, disgust, happiness, and surprise elicit distinct and topographically consistent bodily activation patterns (Nummenmaa et al., 2014; Hietanen et al., 2016; Volynets et al., 2020). For example, anger and fear predominantly activate the upper torso, arms, and head, while sadness is linked to diminished sensations in the limbs. Disgust is consistently localized to the abdomen and throat, possibly reflecting aversive gut reactions and nausea (Nummenmaa et al., 2014). Despite the consistent findings across individuals, individual differences in interoceptive sensitivity (Jung et al., 2017; Tanaka et al., 2024) and neurodivergent conditions such as autism (Palser et al., 2021), exteroceptive sensitivity (Minatoya et al., 2025), or schizophrenia (Torregrossa et al., 2019) shape the granularity, clarity, and extent of emotional bodily maps. Also, children show increasingly differentiated maps with age, indicating a developmental refinement of emotional embodiment (Hietanen et al., 2016). Second, these patterns are not limited to basic emotions. Higher-order and social emotions—such as love (Rinne et al., 2023), pride, guilt, morality (Atari et al., 2020), aesthetics (Daikoku et al., 2024), and empathy (Sachs et al., 2019)—are also reliably mapped onto the body. For instance, different types of love show gradient patterns, from head-centric sensations in abstract forms (e.g., love of wisdom and moral love) to whole-body warmth in romantic love (Rinne et al., 2023). Another study showed that moral emotions exhibit topographically dissociable maps, with violations of fairness or loyalty appearing in the chest, and purity violations centered in the abdomen (Atari et al., 2020). Music-induced emotions similarly generate highly specific mappings: joyful music activates the arms and legs, sad music the chest, and complex experiences such as beauty or surprise manifest in the heart or gut (Putkinen et al., 2024; Daikoku et al., 2024). Third, bodily maps appear to track not only the valence of emotion but also its qualitative "texture". Recent studies have incorporated additional dimensions, such as bodily weight or lightness (Hartmann et al., 2023), to index the subjective density of emotion. Happiness and pride were associated with feelings of lightness throughout the body, while sadness and depression were marked by feelings of heaviness and inertness. These findings complement earlier research showing that individuals with major depressive disorder exhibit overall blunted and undifferentiated bodily maps, suggesting that emotional flattening may manifest as reduced bodily differentiation (Lyons et al., 2021).

Although many bodily maps align with well-established patterns of bottom-up physiological arousal—such as elevated heart rate or gut sensations—not all follow this trajectory. Several maps may instead reflect top-down motor affordances, or even metaphorical associations derived from language and culture. For example, feelings of “cold feet” when afraid or “losing one’s head” when angry might be reported in the absence of measurable physiological change (e.g., no one, after all, literally loses their head), raising questions about the ontological status of what bodily maps represent. Emotional embodiment, then, is neither arbitrary nor purely metaphorical: it is spatially systematic and developmentally emergent—yet also sensitive to individual traits, social context, and semantic framing. In the following section, we propose a theoretical framework grounded in three competing—but complementary—hypotheses regarding the mechanisms underlying these mappings:

1. Body-to-Brain pathway: Bodily maps reflect bottom-up physiological (visceral and autonomic) responses to stimuli, such as increased heart rate by surprisal stimuli.

2. Brain-to-Body pathway: Bodily maps capture top-down behavioral tendencies, like the impulse to dance with excitement.

3. Brain’s conceptual processing: Bodily maps are rooted in metaphorical or conceptual representations of emotion that are projected onto the body, such as “blow one’s head”, even in the absence of measurable physiological change.

These mechanisms are not mutually exclusive; rather, they may interact in dynamic and context-dependent ways to generate the embodied experience of emotion. In particular, the Brain-to-Body pathway should be interpreted as differences in relative effector weighting, not as claims that specific emotions reside in particular body parts. Disentangling their respective contributions can illuminate why bodily maps vary across individuals with different sensory, cultural, or developmental backgrounds, and whether such maps might serve as language-independent tools for assessing emotional and personality traits. This potential is particularly salient in cross-cultural and developmental contexts, where verbal reports are often limited or biased.

**Table 1. Overview of Findings in Body Mapping Study.**

| Area | Emotion/Cognition | Key Insight | Reference |
|---|---|---|---|
| **Head** | General emotions | Reflects facial muscle tension, tear production, skin temperature, and emotion-induced thoughts | Nummenmaa et al., 2014 |
| | Disgust | Sensation in the mouth associated with gastrointestinal activity | Molaeinezhad et al., 2021 |
| | Surprise | Head sensations observed in children as young as six | Hietanen et al., 2016 |
| | | Depression: Head-dominant sensation during musical prediction error linked to negative affect | Tanaka & Daikoku, 2024 |
| | Joy | Head sensations prominent during visual art appreciation across emotions, especially delight | Schino et al., 2024 |
| | Love | All types of love are felt in the head; weaker love (e.g., nature, strangers) remains head-localized | Rinne et al., 2023 |
| | Sociality | Strong head sensations in positive social contexts | Novembre et al., 2019 |
| | Empathy | Head- and chest-centered full-body sensation during empathy for others' joy; weaker in children | Sachs et al., 2019 |
| | Morality | Morally aversive experiences cognitively felt in the head | Atari et al., 2020 |
| | Cognition | Non-emotional cognition (e.g., thinking, recalling) evokes weak non-head sensations | Nummenmaa et al., 2018 |
| **Chest** | Basic emotions | Increased sensations in the upper chest reflecting heart rate and respiration changes | Nummenmaa et al., 2014 |
| | Sadness | Reflects awareness of changes in heart rate and breathing | Putkinen et al., 2024 |
| | | Stronger in women; consistent across cultures | Volynets et al., 2020 |
| | Anger | Stronger in individuals with higher interoceptive accuracy | Jung et al., 2017 |
| | Fear | Distinct chest sensations evoked by fear and anger related to war, climate change, and COVID-19 | Herman et al., 2022 |
| | Pleasure | Stronger in response to hierarchically structured musical sequences | Minatoya et al., 2025 |
| | Happiness | Stronger than head sensations during visual art appreciation | Schino et al., 2024 |
| | | Recognized early in development, despite increasing emotional differentiation with age | Hietanen et al., 2016 |
| | Love | Romantic and sexual love extends sensations from head to chest | Rinne et al., 2023 |
| | Aesthetic | Chest sensations linked to pleasure and aesthetic experience during musical prediction errors | Daikoku et al., 2024 |
| | Creativity | Chest sensations during high uncertainty correlate with creativity ratings | Daikoku et al., 2024 |
| | Sociality | Stronger in positive social situations | Novembre et al., 2019 |
| | Empathy | Head- and chest-centered full-body sensation during empathy for others' joy; weaker in children | Sachs et al., 2019 |
| | Morality | Evoked by violations of harm, fairness, loyalty, and authority | Atari et al., 2020 |
| **Abdomen** | Fear | Stronger abdominal sensations in individuals with high interoceptive sensitivity | Jung et al., 2017 |
| | Disgust | Linked to gastrointestinal and throat-related sensations | Nummenmaa et al., 2014 |
| | | Disgust-abdomen link is culturally universal | Volynets et al., 2020 |
| | | Higher interoceptive sensitivity linked to stronger gut sensations | Jung et al., 2017 |
| | Negative emotions | Link to fear, disgust, sadness, surprise, anxiety, depression, shame, and jealousy | Herman et al., 2022 |
| | Uncertainty | Complex rhythms and unclear keys evoke abdominal sensations | Putkinen et al., 2024 |
| | Disgust | Intestinal sensations linked to gastrointestinal processing | Molaeinezhad et al., 2021 |
| | Surprise | Evoked during expected music; linked to aesthetic experience | Daikoku et al., 2024 |
| | Morality | Moral violations related to purity evoke nausea in the gut | Atari et al., 2020 |
| **Upper limbs** | Anger | Stronger hand sensations in individuals with high interoceptive sensitivity | Jung et al., 2017 |
| | | Reflects preparation for approach-oriented action | Nummenmaa et al., 2014 |
| | | Arm sensations in anger and happiness are culturally universal | Volynets et al., 2020 |
| | | Arm sensations linked to anger during visual art appreciation | Nummenmaa et al., 2023 |
| | Fear | Arm sensations linked to fear during visual art appreciation | Nummenmaa et al., 2023 |
| | Empathy | Arm sensations linked to empathy during visual art appreciation | Nummenmaa et al., 2023 |
| **Lower limbs** | Sadness | Characterized by energy loss and reduced activity | Nummenmaa et al., 2014 |
| | | Sadness-leg sensation link is culturally universal | Volynets et al., 2020 |
| | | Reduced leg activity in adults not observed in young children | Hietanen et al., 2016 |
| | Sadness/Depression | Sadness and depression during COVID-19 linked to reduced leg sensations | Herman et al., 2022 |
| | Morality | General moral violations associated with reduced leg activity | Atari et al., 2020 |
| | Empathy | Reduced limb strength during sadness is more strongly shared in those with high empathy | Sachs et al., 2019 |
| **Limbs** | Sadness | Hand and leg sensations stronger in individuals with high interoceptive sensitivity | Jung et al., 2017 |
| | Happiness | Rhythmic music with clear beat elicits strong pleasant sensations in limbs | Putkinen et al., 2024 |
| | Sociality | Weakened limb sensations in negative social situations | Novembre et al., 2019 |
| | Empathy | Reduced limb strength during sadness is more strongly shared in those with high empathy | Sachs et al., 2019 |
| **Whole body** | General emotions | Depressed individuals on medication show reduced bodily sensations outside head, heart, and gut | Lyons et al., 2021 |
| | Fear | Fear from visual art linked to whole-body sensations | Nummenmaa et al., 2023 |
| | Sadness/Depression | Experienced as bodily heaviness | Hartmann et al., 2023 |
| | Happiness | Linked to whole-body sensations | Nummenmaa et al., 2014 |
| | | Association with whole-body sensation is culturally universal | Volynets et al., 2020 |
| | | Experienced as lightness throughout the body | Hartmann et al., 2023 |
| | Love | Stronger love spreads sensations body-wide; extent scales with intensity | Rinne et al., 2023 |
| | | Secondary emotions like love evoke less bodily sensation than basic emotions | Nummenmaa et al., 2014 |
| | Aesthetic | Aesthetic experience from visual art linked to whole-body sensation | Nummenmaa et al., 2023 |
| | Being moved | Feeling moved during visual art appreciation is linked to whole-body sensation | Nummenmaa et al., 2023 |
| | Empathy | Greater overlap between self and other bodily maps in those with high empathy | Sachs et al., 2019 |
| | | Empathy from visual art linked to whole-body sensation | Nummenmaa et al., 2023 |

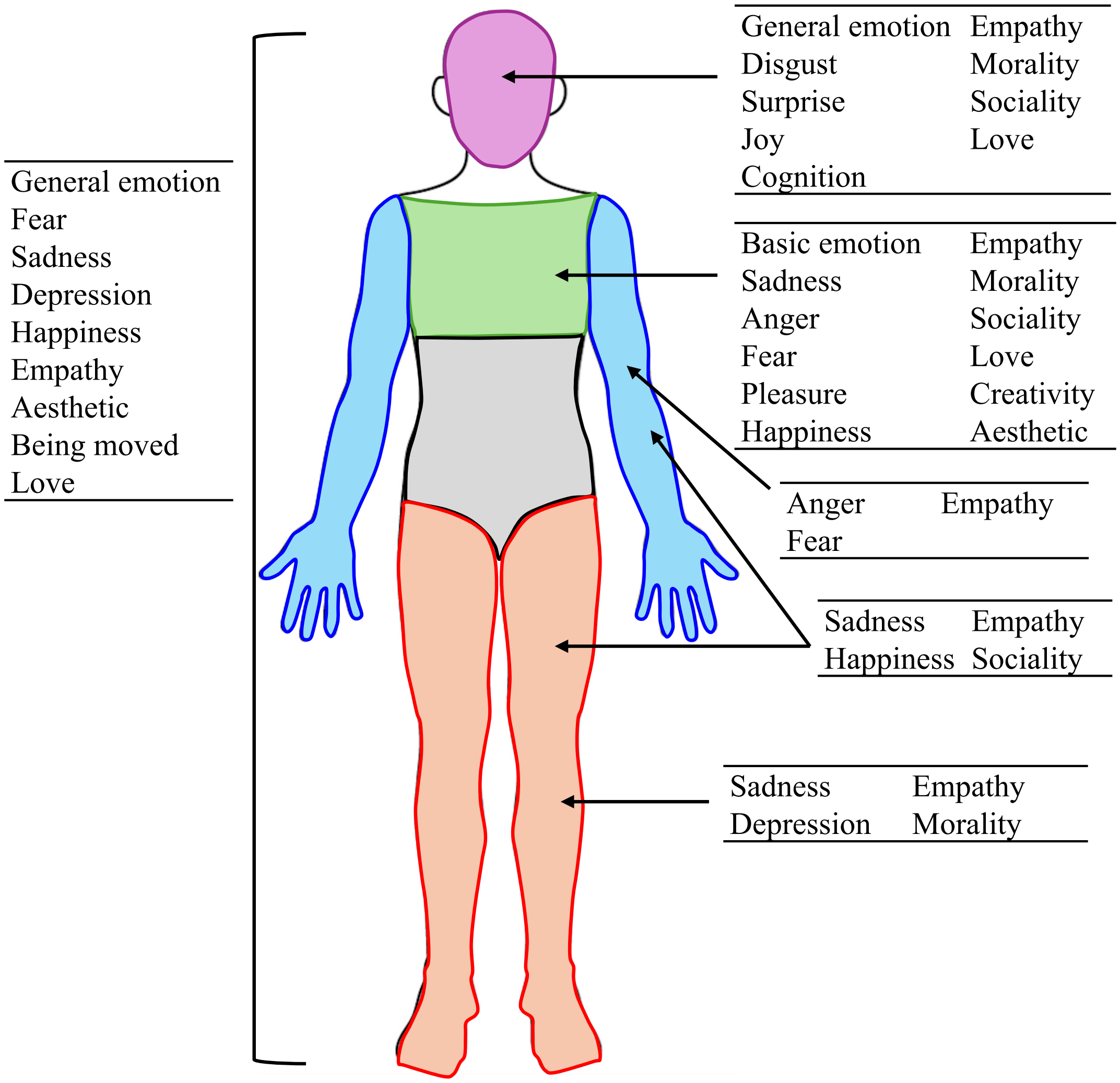

**Figure 1. Annotated body-map of emotions and cognition.** The figure illustrates body regions associated with distinct emotional and cognitive experiences, as summarized in Table 1. This visualization offers an at-a-glance synthesis of the distribution of embodied experiences.

# 3. Key insights into the Emergence of Bodily map

## 3.1. Body-to-Brain pathway: Arising from Physiological Responses

Beyond exteroceptive modalities such as audition or vision, emotional experiences are often registered through interoceptive and proprioceptive pathways. These internal channels give rise to visceral sensations—such as a quickening heartbeat during moments of anxiety, or a spine-tingling shiver when gripped by fear or moved by powerful music (Trost et al., 2017; Mori & Iwanaga, 2017; Koelsch & Jäncke, 2015). Among the various bodily systems implicated in emotional experience, interoception—the brain's ability to perceive internal bodily signals—has emerged as a crucial component of emotional awareness and mental functioning (Critchley et al., 2004). This notion can be traced back to William James's theory of emotion (James, 1884), which posited that emotions are nothing more than the brain's perception of physiological changes: *"We feel sorry because we cry, angry because we strike, afraid because we tremble"*. In this framework, the experience of emotion arises from the brain's interpretation of bodily states.

Regarding this, Damasio (1994) also proposed the somatic marker hypothesis, asserting that distinct bodily states—such as changes in heart rate, muscle tone, or gut sensation—are encoded and reactivated by the brain during emotional evaluation. From this perspective, emotions are the neural mappings of evolutionarily adaptive bodily responses. Indeed, cross-cultural studies have shown that people consistently localize different emotions to particular regions of the body, suggesting that emotion-specific bodily maps may reflect biologically ingrained patterns of interoceptive awareness (Nummenmaa et al., 2014). Physiological evidence supports these claims. Salazar-López et al. (2015) used high-resolution thermography to measure changes in bodily surface temperature as participants experienced various emotions. They found distinct thermal signatures associated with each emotion—for instance, increased warmth in the cheeks and nose during love, and cooling across the face during sadness. These thermographic patterns corresponded closely with participants' self-reported bodily maps, providing converging evidence for the physiological basis of embodied emotional experience. Neuroimaging studies further reveal that emotion-related bodily maps have neural substrates. Nummenmaa et al. (2018) demonstrated that emotions subjectively rated as highly bodily—such as fear or pride—were associated with increased activity in brain regions responsible for interoceptive processing, including the insula and somatosensory cortex. These findings suggest that the subjective experience of emotional bodily sensations is underpinned by the brain's capacity to monitor and simulate internal bodily states. Notably, insular activity correlates with both the intensity and location of subjectively reported sensations, further reinforcing the notion that bodily maps reflect a neurophysiologically grounded phenomenon (Cameron, 2002).

These body–emotion relationships have also been conceptualized through the lens of predictive processing. According to this framework, emotions emerge as the brain attempts to minimize prediction errors between expected and actual bodily inputs (Seth, 2013; Barrett & Simmons, 2015). Emotional arousal may thus result from a mismatch between top-down predictions and bottom-up interoceptive signals. For example, a sudden, unexpected chord in music may trigger a spike in cardiac activity, interpreted by the brain as emotional surprise. The act of breathing deeply in response may then reflect an effort to update internal priors and restore homeostasis (Barrett et al., 2015; Ainley et al., 2016). Recent empirical work supports this account. In a large-scale body mapping study, Daikoku et al. (2024) showed that musical chord progressions characterized by low uncertainty and high surprise elicited subjectively localized sensations in the chest, particularly around the heart, while highly predictable progressions induced sensations predominantly in the abdomen (Table 1). These findings indicate that interoceptive responses to music are shaped by the interplay between prediction error and prior expectation—and that this dynamic is spatially encoded in interoceptive awareness. Intriguingly, cardiac sensations were positively correlated with emotional valence, underscoring the potential of bodily maps as indices of affective processing.

Further evidence for the physiological grounding of bodily maps comes from research on fear processing. Jung and colleagues (2020) demonstrated that heightened bodily states of fear facilitated biased perception of fearful faces, with activity localized to the fronto–insular–temporal network. These findings indicate that interoceptive states are not only subjectively localized as somatotopic bodily sensations but also supported by convergent neural pathways that integrate visceral input with affective evaluation. Together with our example of music-induced chest sensations under conditions of low uncertainty and high surprise (Daikoku et al., 2024), such evidence underscores how bodily maps reflect a dynamic interplay between predictive coding mechanisms and the collective visceral states that shape emotional perception.

Together, these perspectives underscore the notion that emotional experiences are not merely accompanied by bodily changes but are in part constituted by them. Physiological responses—especially those involving interoceptive organs such as the heart and gut—play a crucial role in shaping how emotions are felt, localized, and expressed. While physiological signals provide the substrate for these feelings, they are dynamically shaped by predictive processes. The resulting bodily maps may reflect this interplay.

## 3.2. Brain-to-Body pathway: Emotion-as-action and effector-weighted action maps

Emotions are enacted as much as they are felt. They are communicated through expressive actions—posture, gait, gesture, and facial/vocal behavior—that both shape and are shaped by affective state. Observers reliably infer emotion from whole-body kinematics and postural configurations, with universal regularities and culturally patterned variants (e.g., approach-oriented arm activity in anger; reduced gait energy in sadness; expansive posture in pride) (Atkinson et al., 2004; de Gelder, 2006; Dael et al., 2012; Tracy & Matsumoto, 2008; Jack et al., 2012). On this view, emotional expression is intrinsically action-oriented rather than confined to any single body location.

These action–feeling couplings are supported by well-characterized mechanisms. Emotional contagion—automatic mimicry and synchronization with others' expressions—can induce congruent motor states and corresponding feelings (Hatfield et al., 1993; Niedenthal, 2007). Mirror-neuron–related simulation links action observation to motor representations (Rizzolatti & Craighero, 2004; Gallese & Goldman, 1998). In audition, dorsal auditory–motor pathways allow sounds and music to directly drive motor timing and planning, engaging premotor/supplementary motor areas and basal ganglia during beat perception (Hickok & Poeppel, 2007; Zatorre et al., 2007; Grahn & Brett, 2007). These mechanisms explain why rhythm readily elicits head-nodding or foot-tapping and why one's own or others' actions can transform how emotions are felt in the body.

In this framework, we interpret the topographies observed under the Brain-to-Body pathway as effector-weighted action maps: predictive control policies that allocate motoric gain across effectors as a function of affect and task. This reconciles two facts: (i) the whole body participates in emotional expression, and (ii) regional regularities appear in mapping tasks. The claim is not that an emotion resides in one body part, but that different emotions bias characteristic, context-sensitive distributions—for example, approach-ready arms/hands in anger and happiness, reduced lower-limb drive in sadness, and protective contraction of the torso in fear—consistent with the regularities summarized in Table 1.

While emotional feelings can arise from bottom-up physiological responses, they also emerge through top-down processes—particularly those involving action and motor behavior generated by the brain. Emotions are not solely internal states; they are often enacted through spontaneous and expressive bodily movements. The raised arms of elation, the hunched shoulders of shame, or the averted gaze of guilt all exemplify how emotional experience is embedded in—and shaped by—sensorimotor expression. These movements are not merely outputs of emotional states; they are integral to how emotions are felt and interpreted within the body.

Crucially, some actions play a regulatory role in modulating emotional arousal. The act of taking a deep breath in moments of anxiety or irritation, for instance, can be understood as a homeostatic strategy: a top-down behavioral intervention aimed at rebalancing heightened physiological states. Within the framework of brain's prediction, this action can be interpreted as "interoceptive active inference" (Seth, 2013). Experimental work by Philippot et al. (2002) has demonstrated that breathing patterns themselves can induce emotional states, with slow, deep breathing associated with calmness and rapid, shallow breathing with anxiety. Such findings highlight the pathway from bodily action to bodily sensation, and the emotional state, underscoring that bodily sensations are not merely passive consequences of emotion but can also serve as active tools for its regulation.

This dynamic interplay between prediction, motor action, and emotion is evident in the domain of music. Musical experience is not limited to passive auditory processing; rather, it frequently engages listeners in active, embodied responses that align with rhythm and meter. Rhythmic syncopation—violations of temporal expectation—can trigger spontaneous motor behaviors such as head-nodding or foot-tapping, reflecting the brain's drive to reduce prediction error by adjusting internal timing models to match external rhythmic patterns. Notably, recent body mapping studies have also demonstrated that rhythmically engaging music tends to elicit localized sensations in the legs and feet, further suggesting that musical rhythms are embodied not just cognitively but sensorimotorically (Putkinen et al., 2024). These (even imaginary) motor reactions are thought to reflect the brain's attempts to minimize prediction error by aligning its internal timing models with external musical structure—a process known as music-based active inference or music-based on mental active inference (Koelsch et al, 2019; Vuust et al., 2022). Supporting this interpretation, neuroimaging evidence indicates that listening to moderately syncopated music activates motor regions of the brain, including premotor and supplementary motor areas, even in the absence of overt movement (Matthews et al., 2020). In this way, bodily movement becomes a mechanism through which the brain enacts its predictions and refines emotional engagement with music.

Such examples illuminate a central insight: bodily sensations can emerge not only from bottom-up physiological input but from top-down behavior and actions. This perspective aligns with the "*conceptual act theory*" proposed by Barrett (2017), which posits that the brain does not simply register emotional states from raw interoceptive data but constructs them through predictive, top-down processes such as interoceptive active inference (Seth, 2013). According to this view, emotions arise when the brain applies learned emotion concepts to ambiguous bodily sensations, drawing on past experiences. Even in the absence of marked physiological change, the brain may simulate the expected bodily signature of an emotion—for example, tightness in the head or trembling hands in the case of anger—based on a conceptual template.

In this framework, bodily maps are not straightforward reflections of biological events but are also partly cognitively constructed—subjective sensorimotor patterns generated by the brain to match the meaning of a given emotional state. Importantly, these simulations may arise even when physiological inputs are minimal or absent. This helps explain empirical findings that challenge the notion of fixed physiological "fingerprints" for each emotion. For example, Siegel et al. (2018) showed that autonomic measures such as heart rate, blood pressure, and electrodermal activity substantially overlap across emotional categories, making it difficult to clearly distinguish emotions based on physiology alone. Yet people reliably report distinct bodily maps for different emotions, suggesting that what they feel in the body may reflect conceptually driven expectations as much as physiological reality. From this perspective, even William James's famous proposal—that we do not cry because we are sad, but are sad because we cry—can be reinterpreted within a top-down framework. The initial impulse to cry, as a behavioral act, may trigger a cascade of bodily sensations around the eyes and face. These sensations are then interpreted by the brain as corresponding to sadness. Thus, bodily maps in this model may not arise from physiology alone but from the interplay between action, sensation, and conceptual labeling.

Taken together, these findings suggest that emotional embodiment is shaped by a continuous feedback loop between brain and body, where prediction, movement, and interpretation co-construct the feeling of emotion. Bodily maps, in this view, capture more than the body's reaction—they reflect how the brain organizes emotional experience through action and expectation. Understanding these top-down influences is essential for a complete account of how emotions are felt, enacted, and remembered in the body.

## 3.3. Brain's Conceptual Construction: Arising from Metaphorical Schemas

Emotions are not only grounded in physiological and behavioral responses—they are also imagined, conceptualized, and narratively constructed through the interpretive lens of the mind. While bodily responses provide a foundational substrate, the subjective experience of emotion is profoundly shaped by cognitive processes that impose symbolic structure and metaphorical meaning onto the body. From this perspective, emotional embodiment is not merely a biological phenomenon, but also a conceptual construction—rooted as much in mental representations as in physical sensation.

Language offers one of the clearest portals into this conceptual dimension. Across cultures, idiomatic expressions consistently link emotions to specific parts of the body— "losing one's head" in anger or "cold feet" in fear. Such expressions suggest that bodily sensations may not simply arise from internal signals but from culturally shared metaphors that structure how people expect to feel emotions in their bodies. Importantly, these metaphors are not merely linguistic embellishments; they shape perception itself. This view is supported by the embodiment theory of emotion concepts proposed by Niedenthal and colleagues (2005; 2007), who argued that people partially simulate past bodily states to understand and categorize emotional experiences. For instance, understanding the concept of anger may involve a reactivation of muscle tension or facial heat associated with previous episodes of anger. In this view, the bodily map is not merely a reflection of physiological status, but a re-enactment of prototypical bodily states stored in memory.

These mental simulations frequently draw on culturally embedded metaphorical systems. Kövecses (2000) and Lakoff & Johnson (1980) have shown that across diverse languages, emotions are systematically linked with embodied spatial metaphors—anger is "hot" or "boiling," sadness is "heavy" or "low," and happiness is "light" or "uplifting." Such metaphors not only inform how people talk about emotions but also shape how they localize them within the body. A compelling example is the Japanese expression "*hara-ochi*" (literally, "it drops into the belly"), which is used to describe a moment of emotional resolution or cognitive insight. This association between understanding and abdominal sensation points to a culturally shared mapping of affective meaning onto bodily space. Empirical studies lend credence to these metaphorical associations. In a large-scale body mapping study of musical emotion, Daikoku et al. (2024) found that musically predictable progressions—those that fulfilled the listener's expectations—elicited sensations localized in the abdomen. Participants reported this sensation as meaningful and emotionally satisfying, possibly echoing idiomatic expressions like "it drops into the belly" or "it sits right in the gut". Similarly, more recent work (Daikoku et al., 2023) showed that auditory pitch can influence the spatial distribution of bodily maps: high-pitched sounds were associated with sensations in the upper body, while low-pitched tones were linked with the lower body. These mappings likely do not reflect direct physiological pathways, but rather symbolic spatial schemas grounded in perceptual metaphors of "high" and "low."

These culturally shaped metaphors are not only acquired—they may also become internalized. Waggoner (2010) found that both children and adults consistently used metaphors involving temperature and weight to describe their emotional experiences—for instance, anger as a rush of heat to the head, or sadness as a heavy, sinking feeling. Notably, these metaphoric patterns were consistent across ages, suggesting that cultural learning of emotional metaphors begins early in life. Even when no actual physiological change occurs, individuals may report feeling their "head heat up" when angry—a form of self-fulfilling bodily expectation shaped by culturally acquired schemas. In this sense, bodily maps are not necessarily biological records of change, but metaphorical landscapes imagined and reinforced through language, learning, and culture.

Attention and imagination play an important role in how we feel emotions in the body. For example, when people feel fear or anxiety, they often report strong sensations in the chest or stomach—even if there are no clear physiological changes in those areas. This suggests that where we focus our attention can make certain body parts feel more intense, shaping a bodily map based more on perception than on actual bodily changes. Classic research by Schachter and Singer (1962) showed that simply paying attention to bodily responses can influence how we interpret emotions. Recent studies have found that people with high interoceptive anxiety tend to feel stronger sensations in the chest and heart area (Ehlers & Breuer, 1996; Domschke et al., 2010; Tanaka et al., 2024b), further highlighting the role of focused attention in shaping bodily emotional experiences.

Social context further amplifies these effects. Just as facial expressions and gestures function as tools for emotional communication, descriptions of internal bodily states may also serve a communicative purpose. For instance, people often describe sadness as "pain in the chest"—not only because they truly feel these sensations, but because such expressions align with culturally shared emotional scripts (Shaver et al., 1987). These scripts, often activated unconsciously, help shape how people perceive and report their bodily feelings. In cultures where people learn that "anger is something you feel in your gut," this schema may guide both attention and interpretation, contributing to the formation of consistent bodily maps. In this way, social conventions help scaffold emotional experience, reinforcing culturally patterned bodily maps even in the absence of distinct physiological changes.

In sum, bodily maps are not purely reflections of internal physiological states; they are also shaped by imagination, language, and culture. They offer a canvas upon which people project emotional meaning, often guided by metaphorical, symbolic, and socially shared constructs. By studying how the mind maps emotion onto the body, we can better understand not only the subjective texture of emotional experience but also the shared scaffolding through which it is structured and communicated.

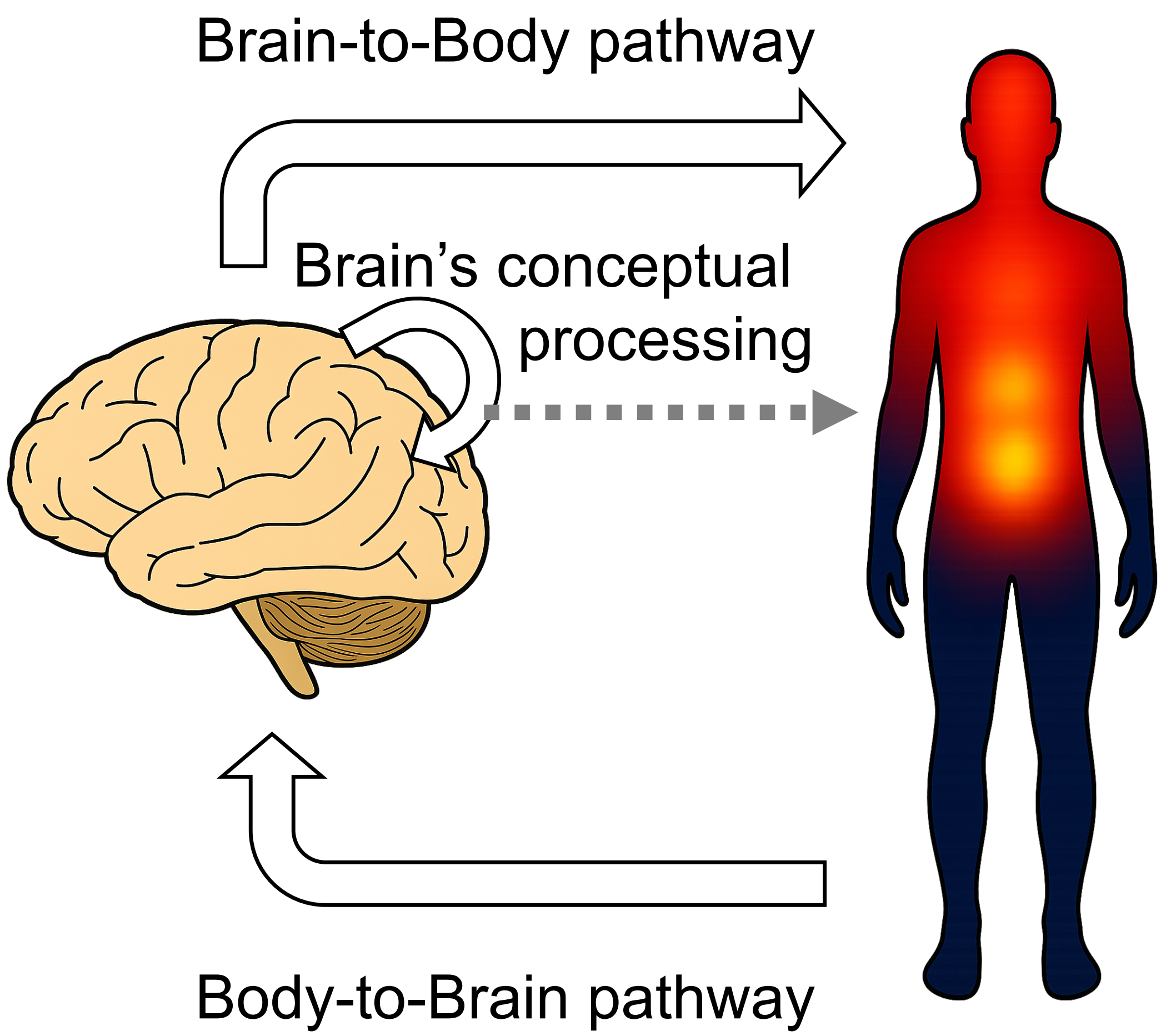

**Figure 2. A Tripartite Framework for the Emergence of Bodily maps.**

# 4. Discussion

Emotional experiences are not disembodied abstractions—they are felt, lived, and mapped within our own body. In this review, we proposed a tripartite framework to account for the emergence of these embodied feelings as captured by bodily maps. This framework includes, body-to-brain pathway, suggesting that bodily maps arise from bottom-up interoceptive and autonomic signals; the brain-to-body pathway, which emphasizes top-down modulation through motor behaviors and action tendencies; and the conceptual construction, positing that bodily sensations may be constructed through metaphorical schemas, even in the absence of physiological reactions and motor behaviors. These mechanisms are not mutually exclusive. Rather, they should be dynamically shaped by context, culture, development, and individual differences. Here, we discuss how these three pathways interact—at times reinforcing one another, and at times offering distinct or even conflicting explanations for emotional embodiment.

## 4.1. Physiological and Behavioral Origins vs. Conceptual Construction

Support for the physiological view comes from research demonstrating a correspondence between emotional experiences, interoceptive awareness, and bodily self-report. For example, Nummenmaa and colleagues (2014) suggested that subjective bodily maps reflect the integrated perception of multiple concurrent changes—such as elevated heart rate, muscle tension, and increased body temperature during anger, which together give rise to the sensation of “heat in the chest or head.” Importantly, individuals with higher interoceptive sensitivity tend to report more vivid and differentiated bodily maps when experiencing emotions, implying that real bodily signals contribute substantially to the formation of these maps (Jung et al., 2017). Neuroimaging findings further support this view. In a large-scale analysis of emotion-related bodily maps, Nummenmaa et al. (2018) demonstrated that emotions rated as highly bodily were associated with increased activation in brain regions implicated in interoception and bodily monitoring, including the insula and somatosensory cortex. These regions are known to encode the intensity and spatial characteristics of internal bodily states (Cameron, 2002), suggesting that bodily maps may indeed be neurophysiologically grounded representations of affective bodily experience.

Yet, meta-analytic findings by Siegel et al. (2018) reveal that autonomic markers such as heart rate, blood pressure, and electrodermal activity show substantial overlap across emotional categories, making it difficult to identify distinct physiological “fingerprints” for each emotion. Despite this physiological ambiguity, individuals consistently report clearly differentiated bodily maps for anger, sadness, fear, and other emotions. This suggests that subjective bodily maps may not mirror physiological changes directly, but are instead shaped by how the brain conceptually interprets, categorizes, and labels bodily sensations. Studies in clinical populations further illustrate this dissociation. Palser et al. (2021) found that autistic children exhibited less differentiated emotional bodily maps, despite showing no significant differences in objective interoceptive sensitivity compared to typically developing peers. Similarly, individuals with schizophrenia may display altered bodily maps not because of reduced physiological input, but due to impaired associations between bodily sensations and emotional concepts (Torregrossa et al., 2019). These findings indicate that the construction of bodily maps involves not just the sensing of bodily states, but also their interpretation through learned emotional frameworks.

Yet, the intriguing phenomenon arises when emotions are felt in the body in the absence of any clear physiological signals. There are numerous instances in which people localize feelings to specific body parts without corresponding somatic changes. Take, for example, the Japanese expression "funi-ochiru", literally meaning “it drops into the gut,” which refers to a moment of cognitive and emotional resolution. There is, of course, no actual sensation of something physically dropping into the gut or abdomen. Such an example highlights a fundamental question: Why do humans consistently use specific body parts to represent emotional experience, even when no underlying physiological change or motor behavior can be detected? If these sensations do not originate from the body, why do we experience them as being in the body—and why in those particular locations? Addressing this question requires us to look beyond the biological substrate and into the symbolic, cultural, and conceptual scaffolding through which emotional experience is structured. It is here that metaphor, language, and shared cognitive schemas may play a crucial role in shaping how emotions are embodied—not as raw sensory data, but as meaningful, spatially anchored constructs of the mind.

These distinctions between physiological, behavioral, and conceptual sources of embodiment become especially salient when examining emotion-specific bodily regions across different types of stimuli. For instance, rhythmically engaging music has been shown to consistently activate sensations in the legs and feet—regions typically associated with movement and motor readiness—suggesting a behavioral origin grounded in embodied action tendencies (Putkinen et al., 2024). Here, the bodily map reflects not internal physiological changes but the impulse to move, consistent with the brain-to-body pathway. By contrast, surprising musical elements—such as sudden chord shifts or unexpected modulations—often elicit sensations in the chest or heart, pointing to physiological origin and body-to-brain pathway (Daikoku et al., 2024). This sensation appears to reflect the brain’s prediction error, in line with the interoceptive predictive coding (Barrett & Simmons, 2015; Seth, 2013).

Other emotions and body regions, however, seem to be shaped more by conceptual construction than by physiology or behavior. For example, aesthetically pleasing but highly predictable musical stimuli elicit localized sensations in the abdomen among Japanese listeners, often accompanied by a feeling of resolution or insight (Daikoku et al., 2024). This is consistent with the cultural metaphor of funi ochiru (“it drops into the gut”) and suggests that such bodily maps are, possibly, mediated by conceptual schemas rather than somatic signals. Similarly, a layered interpretation of anger expressions in Japanese: atama ni kuru (“goes to the head”) reflects a cognitively driven representation likely associated with prefrontal activation; mukatsuku ("nauseating in the chest and abdomen"), aligning with visceral changes and elevated arousal; whereas hara ga tatsu ("the belly rises") appears to be a culturally learned, metaphorically anchored construct with minimal physiological basis. These examples collectively demonstrate that emotional localization in bodily maps cannot be attributed to a single source but instead emerges from a confluence of interoceptive input, behavioral expression, and conceptual understanding, modulated by cultural and contextual factors.

Across these diverse approaches, several consistent patterns emerge that allow a clearer synthesis of the literature. Basic emotions such as fear and disgust reliably involve chest and abdominal/gastrointestinal sensations, while sadness is often localized in the chest and lower limbs, and happiness extends to the whole body with feelings of lightness (Nummenmaa et al., 2014; Hartmann et al., 2023; Volynets et al., 2020). Disgust in particular shows strong convergence across cultures in gut- and throat-related sensations (Volynets et al., 2020; Molaeinezhad et al., 2021). Music studies demonstrated how prediction-related mechanisms drive sensations specifically in chest and abdomen (Daikoku et al., 2024; Putkinen et al., 2024), while metaphorical language studies highlight culturally specific mappings—for instance, anger being felt in the chest in many Western contexts but in the abdomen (hara) in Japanese expressions (Kövecses, 2000; Wierzbicka, 1999). Despite methodological diversity (silhouette drawings, click-based music paradigms, linguistic analyses), these strands converge on the view that bodily maps are structured, recurrent, and partly culture-universal, while still sensitive to local schemas. This synthesis underscores that body mapping is not merely descriptive but reveals systematic correspondences between emotion types and body areas, both within and across cultures.

## 4.2. Modality and Multimodal Integration of Bodily Maps

Evidence suggests that the topography of felt emotion depends, in part, on the inducing modality. Visual-art studies indicate prominent head- and chest-related feelings, along with limb involvement and whole-body states during awe, fear, empathy, and "being moved" (Nummenmaa & Hari, 2023; Schino et al., 2024). In our summary table, visual art-elicited joy shows pronounced head sensations, whereas happiness can be stronger in the chest than the head; fear and empathy can involve the arms, and aesthetic experience often generalizes to whole-body feelings. By contrast, auditory studies highlight different biases: rhythmically engaging music evokes pleasant sensations in the limbs, while predictive dynamics in musical harmony drive chest-centered feelings to surprising events and abdomen-centered feelings to highly predictable progressions (Putkinen et al., 2024; Daikoku et al., 2024). Together, these patterns may imply that visual inputs may weight perceptual–semantic and postural/defensive tendencies (head, chest, arms, whole-body), whereas auditory inputs more strongly engage motor affordances and cardiorespiratory/gastrointestinal interoception (limbs, chest, abdomen).

A mechanistic account follows from predictive processing: each modality furnishes distinct error signals and action affordances that bias where sensations are felt. Visual scenes often recruit appraisal, vigilance, and approach–avoidance schemas (head/face, chest, arms), whereas musical structure recruits timing/motor predictions (limbs) and visceral responses to expectation violation or confirmation (chest/abdomen). Despite these biases, convergence is expected at supramodal interoceptive hubs (e.g., insula, somatosensory cortex) that track intensity and location of felt states across modalities, consistent with prior works linking "highly bodily" emotions to interoceptive cortices. This framework helps explain both modality-specific topographies and cross-modal commonalities (e.g., whole-body "being moved" by either art or music).

We propose within-subject designs that (i) orthogonally manipulate Modality (audio-only, visual-only, audiovisual), (ii) contrast congruent (e.g., fearful image + tense music) versus incongruent pairings (e.g., fearful image + joyful music) to test precision-weighted cue combination, and (iii) use time-resolved body-mapping (e.g., click-based responses within seconds of onset) to capture evolving topographies. Concurrent autonomic and motor measures (e.g., ECG/respiration/EDA, EMG of distal limbs) would adjudicate whether limb- or chest-dominant maps arise from modality-specific physiology versus top-down simulation.

## 4.3. Embodied Conceptual Metaphor underlying Body Mapping

Conceptual Metaphor Theory (Lakoff & Johnson, 1980; Kövecses, 2000) proposes that abstract emotional experiences are systematically structured through bodily and spatial metaphors grounded in sensorimotor interactions with the physical world. Emotions are not only felt in the body but also understood through the body. For instance, the metaphor "understanding is digestion" is embedded in the Japanese phrase funi ochiru, meaning "it drops into the gut," which signifies sudden comprehension. Although no actual visceral sensation occurs—nothing is literally dropping into the gut—the expression conveys the cognitive event through a deeply embodied schema, suggesting that the process of making sense is felt as internalization (Johnson, 1987). Similar metaphorical scaffolding underlies emotional expressions across languages. Anger, for example, is frequently conceptualized as a hot fluid in a sealed container, where pressure builds until it bursts— "boiling with anger" in English, or hara ga tatsu (lit. "the belly rises") in Japanese (Lakoff, 1987). In this model, the body serves as both the vessel and the expressive medium of emotional escalation.

Yet while such metaphors are structurally similar across cultures, their bodily localizations vary significantly. The ANGER IS HEAT schema appears to be cross-culturally recurrent (Kövecses, 2005), but the spatial focus of that heat differs. In English and many Indo-European languages, anger tends to be situated in the head or chest—expressed through phrases such as "hot-headed" or "heart pounding with rage." By contrast, Japanese anger metaphors trace a vertical trajectory through the body, beginning in the abdomen (hara), rising through the chest, and culminating in the head (Matsuki, 1995; Kövecses, 2000). These metaphors reflect deep-seated cultural schemas in which the hara is seen as the locus of sincerity, emotional truth, and internal will (Wierzbicka, 1999; Ma-Kellams, 2014).

Such cultural variation in emotional embodiment suggests that bodily maps are shaped not only by direct somatic experience but also by metaphorical and symbolic systems. These systems—shared within a linguistic and cultural community—provide a cognitive framework that guides the interpretation of diffuse physiological cues. For example, the widespread agreement among Japanese speakers that “anger resides in the belly” likely reflects not just universal autonomic signals but also culturally acquired metaphorical mappings. This is supported by evidence showing that even in the absence of distinct physiological signatures, individuals report stable, differentiated emotional bodily maps that align with culturally specific metaphors (Siegel et al., 2018; Torregrossa et al., 2019).

Intercultural differences in interoceptive focus further reinforce this interpretation. Research suggests that East Asian cultures—including Japan and China—endorse more holistic, embodied understandings of emotion (Wierzbicka, 1999). In these traditions, emotions are tightly interwoven with internal bodily states, a perspective reflected in both language and traditional medicine. For instance, Traditional Chinese Medicine views emotions and physiology as dynamically integrated via the circulation of vital energies such as qi (Yu, 2002). This conceptual grounding manifests in frequent references to gut, chest, and bodily sensations in everyday emotion talk.

Building on prior research on emotional bodily maps, we hypothesize that different metaphorical expressions—such as those used to describe anger in Japanese—reflect distinct layers of emotional processing: atama ni kuru (“goes to the head”) may correspond to cognitively mediated anger; mukatsuku (“nauseating”) may reflect visceral, affectively driven anger associated with physiological changes in the heart and abdomen; and hara ga tatsu (“the belly rises”) may represent a culturally specific, metaphorically grounded expression. This view aligns with previous neurophysiological research showing that cognitively oriented terms like “thinking” and “remembering” are predominantly associated with head-based bodily sensations (Table 1, Nummenmaa et al., 2018), whereas individuals with higher interoceptive sensitivity tend to report stronger chest sensations when experiencing anger—suggesting that chest-based anger representations may be more tightly coupled with physiological arousal (Jung et al., 2017). In contrast, the mapping of anger to the abdomen may be less dependent on direct interoceptive feedback and more reflective of culturally embedded metaphorical schemas.

In sum, while physiological and behavioral mechanisms provide a foundational substrate for emotional bodily maps, embodied conceptual metaphors play a pivotal role in shaping how emotions are subjectively localized and experienced. Emotional bodily maps, then, are not simply representation of internal states—they are interpretations filtered through shared bodily models, linguistic metaphors, and cultural knowledge. This interplay between body and concept not only helps explain cultural variation in emotional embodiment but also clarifies why emotions can feel so vividly "bodily" even when no measurable somatic changes occur. Future research would benefit from further integrating conceptual metaphor theory with empirical investigations of interoception, emotion, and culture, offering a more comprehensive account of the embodied mind.

# 5. Conclusion Remarks

This review has examined the emergence of emotional bodily maps by integrating empirical evidence, theoretical frameworks, and cross-cultural perspectives. We proposed a tripartite framework for understanding the origins of these maps, emphasizing the interplay among physiological input, behavioral action, and conceptual construction. Emotions are not confined to the brain or viscera alone—they are lived through the body, shaped by our actions, and structured through culturally acquired schemas. Body-to-brain pathways provide the physiological foundation for emotional sensation, particularly through interoceptive organs such as the heart, lungs, and gut. Yet brain-to-body pathways—manifested in expressive movements like foot-tapping to rhythmic music—also play a crucial role in generating and amplifying emotional experience. Beyond these somatic and motoric foundations, conceptual metaphors guide how sensations are interpreted, categorized, and expressed. Converging evidence from recent work suggests that different emotions and body regions align with different origins: rhythmically induced sensations in the feet emerge from behavioral engagement; cardiac responses to musical surprise from physiological arousal; and abdominal feelings of understanding or insight from culturally embedded conceptual metaphors. Some of the most intense emotional experiences—such as being "moved beyond words" by awe-inspiring natural beauty or moments of profound moral clarity—are difficult to verbalize or measure, yet are nonetheless deeply and specifically localized in the body. These experiences illustrate why bodily maps are so vital: they offer a means of capturing emotions that escape traditional linguistic or physiological frameworks. In such cases, the body speaks where words fall short.

Importantly, body mapping may offer a language-independent tool for assessing emotional traits and capacities. Because bodily maps rely on introspective reporting of felt sensations rather than verbal articulation, they may be especially useful in settings where language is limited, unreliable, or biased—such as in cross-cultural comparisons, developmental studies with young children, or clinical contexts involving individuals with language or communication difficulties. By clarifying how the three pathways—physiological, behavioral, and conceptual—contribute to the embodied experience of emotion, bodily maps may open new avenues for capturing personal and cultural variation in emotion beyond the constraints of language. Ultimately, bodily maps make visible what often remains invisible: the internal geography of feeling. They allow us to trace the outlines of joy, fear, sadness, or awe as they unfold across the flesh. And in doing so, they bring us closer to a holistic understanding of emotion—not merely as something we speak, but as something we deeply feel in our body.

## Competing Interests

The authors declare no competing financial interests.

## Author Contributions

T.D. wrote the draft of the manuscript and figure. T.D. M.M., and M.T. edited and finalized the manuscript.

## Acknowledgements

This research was supported by the JSPS KAKENHI (25H0048, 25K03206, 24K22302, 24H01539), Japan. The funding sources had no role in the decision to publish or prepare the manuscript.

## Data Availability

All the data files are available from the manuscript.